\pdfoutput=1
\documentclass[12pt]{article}\usepackage{qoper}
\begin{document}

\title{QOperAv, a Code Generator\\
for Generating Quantum Circuits
for Evaluating Certain Quantum Operator Averages}

\author{Robert R. Tucci\\
        P.O. Box 226\\
        Bedford,  MA   01730\\
        tucci@ar-tiste.com}

\date{ \today}

\maketitle

\vskip2cm
\section*{Abstract}

This paper introduces
QOperAv v1.5,
a Java application
available for free. (Source code
included in the distribution.)
QOperAv is a ``code generator"
for generating quantum circuits.
The quantum
circuits generated by QOperAv
can be used to evaluate
with polynomial efficiency the
average of $f(A)$
for some simple (that is, computable with polynomial efficiency) function $f$ and
a Hermitian operator
$A$, provided that we know
how to compile $\exp(iA)$
with polynomial efficiency.
QOperAv
implements an algorithm
described in earlier papers,
that combines
various standard techniques such as
quantum phase estimation
and quantum multiplexors.

\newpage

\section{Introduction}

We say a unitary operator
acting an array of qubits has been
compiled if
it has been expressed
as a Sequence of Elementary
Operations (SEO), where by elementary
operations we mean few-qubit
(like 1 and 2-qubit) operations such
as CNOTs and single-qubit rotations.
SEO's are often represented as quantum circuits.

This paper introduces\footnote{The reason
for releasing the first public
version of QOperAv with such an odd
version number is that QOperAv shares
many Java classes with other previous Java
applications of mine
(QuanSuite
discussed in Refs.\cite{quantree,quanfou,quanfruit},
QuSAnn
discussed in Ref.\cite{TucQusann},
Multiplexor Expander
discussed in Ref.\cite{TucQusann},
and Quibbs discussed in Ref.\cite{TucQuibbs}),
so I have made the
decision to give all these applications
a single unified version
number.}
QOperAv v1.5,
a Java application
available for free. (Source code
included in the distribution.)
The name ``QOperAv" is an abbreviation of
the phrase ``Quantum Operator Average",
and is
pronounced like the street, ``Copper Av.".
QOperAv is a ``code generator"
for generating quantum circuits.
The quantum
circuits generated by QOperAv
can be used to evaluate
with polynomial efficiency the
average of $f(A)$
for some simple (that is, computable with polynomial efficiency) function $f$ and
a Hermitian operator
$A$, provided that we know
how to compile $\exp(iA)$
with polynomial efficiency.
Such averages arise, for example, in
an
algorithm
described in Ref.\cite{TucZ} for evaluating
partition functions with a quantum computer.

Apart from its usefulness
as a code generator, QOperAv
is interesting in that it required
very few lines of code to write, because
it relies on classes that
form part of a large class library
that had been written previously. This class library
has been  used previously to construct
many other applications (for example,
QuanSuite,
QuSAnn,
Multiplexor Expander,
and Quibbs).

QOperAv
implements an algorithm
discussed in Ref.
\cite{TucZ}.
The quantum circuit
generated by QOperAv
includes some quantum multiplexors.
The Java application Multiplexor Expander
(see Ref.\cite{TucQusann})
allows the user to replace
each of those multiplexors by
a sequence of more elementary
gates such as multiply
controlled NOTs and qubit rotations.
Multiplexor Expander is also
available for free, including source code.

\section{Input Parameters}

The quantum circuit generated
by QOperAv is described in detail in
Ref.\cite{TucZ}.
Using the notation
of Ref.\cite{TucZ}, the
circuit
depends on the following inputs:

\begin{description}
\item[$\nb$:] This is a positive integer.

\item[$\nb_\rvj$:] This is a positive integer.

\item[$\gamma$:] This is a positive
real.

\item[$\Delta t$:] This is a positive real.

\item[for $p=0, 1, 2, \ldots, \nb_\rvj-1$,
a quantum circuit for
$\exp(i 2^p A\Delta t)$:]
We call the unitary operator $\exp(iA\Delta t)$
 an ``atom"
and  the $\nb$ qubits
it acts on, the atom qubits.
The
demonstration version of QOperAv uses
as an atom the
circuit for an $\nb$-qubit quantum Fourier
transform,
and it raises the atom to the $2^p$-th
power by placing
 the atom inside a LOOP that repeats
 $2^p$ times,
but both
this particular atom
and this method of
raising the atom to a power
can be changed easily by
subclassing the class of
QOperAv that defines this.
In particular,
rather than raising
the atom to a power by
repeating the atom circuit,
the user could
raise the atom to the $2^p$-th power
by replacing the parameter
$\Delta t$ by
$2^p\Delta t$ in the atom circuit.

\item[ a quantum circuit for $V$:]
The unitary operator $V$
acts on the $\nb$ atom qubits.
The
demonstration version of QOperAv uses
for $V$ the
circuit for an $\nb$-qubit quantum Fourier
transform, but
this can be changed easily by
subclassing the class of
QOperAv that defines this.

\item[function $f:\RR\rarrow \RR$:]
 The
demonstration version of QOperAv uses
$f(\xi) = e^{-(0.1)\xi}$, but
this can be changed easily by
subclassing the class of
QOperAv that defines this.

\end{description}

Let $\ns = 2^\nb$ and
$\ns_\rvj = 2^{\nb_\rvj}$.
The Hermitian operator $A$ is assumed to have
non-negative eigenvalues.
Furthermore, $\Delta t$
is assumed to be small enough
that

\beq
A_x \frac{\Delta t}{2\pi}<
 \frac{\ns_\rvj-1}{\ns_\rvj}
 \;
 \eeq
for all eigenvalues $A_x$ of $A$.
Furthermore, we assume that

\beq
0\leq\gamma
f(\frac
{2\pi \;j}
{\Delta t \;N_{S\rvj}}
)\leq 1
\;
\eeq
for $j=0, 1, 2, \ldots \ns_\rvj-1$.

\section{Output Files}

QOperAv outputs 3 types of files: a Log File,
an English File and a Picture File.

A Log File
records all the input and output
parameters that the user entered into the
{\bf Control Panel} (see Sec.\ref{sec-control}),
 so
the user won't forget them.

An English File gives
an ``in English" description
of a quantum circuit.
It
completely specifies the output SEO.
Each line in it
represents one elementary operation,
and time increases as we move downwards
in the file.

A Picture File
partially specifies the output SEO.
It gives an ASCII picture of
the quantum circuit.
Each line in it represents
one elementary operation,
and time increases as we move downwards
in the file.
There is a one-to-one onto correspondence
between the rows of corresponding English
and Picture Files.

English and Picture Files are
used in many of my previous computer programs.
I've explained those files
 in detail in previous
papers
so I won't do so again here. See,
for example, Ref.\cite{TucQuibbs} for
a detailed description of the
content of those
files and how to interpret that content.

\section{Control Window}
\label{sec-control}

\begin{figure}[h]
    \begin{center}
    \includegraphics[scale=.70]{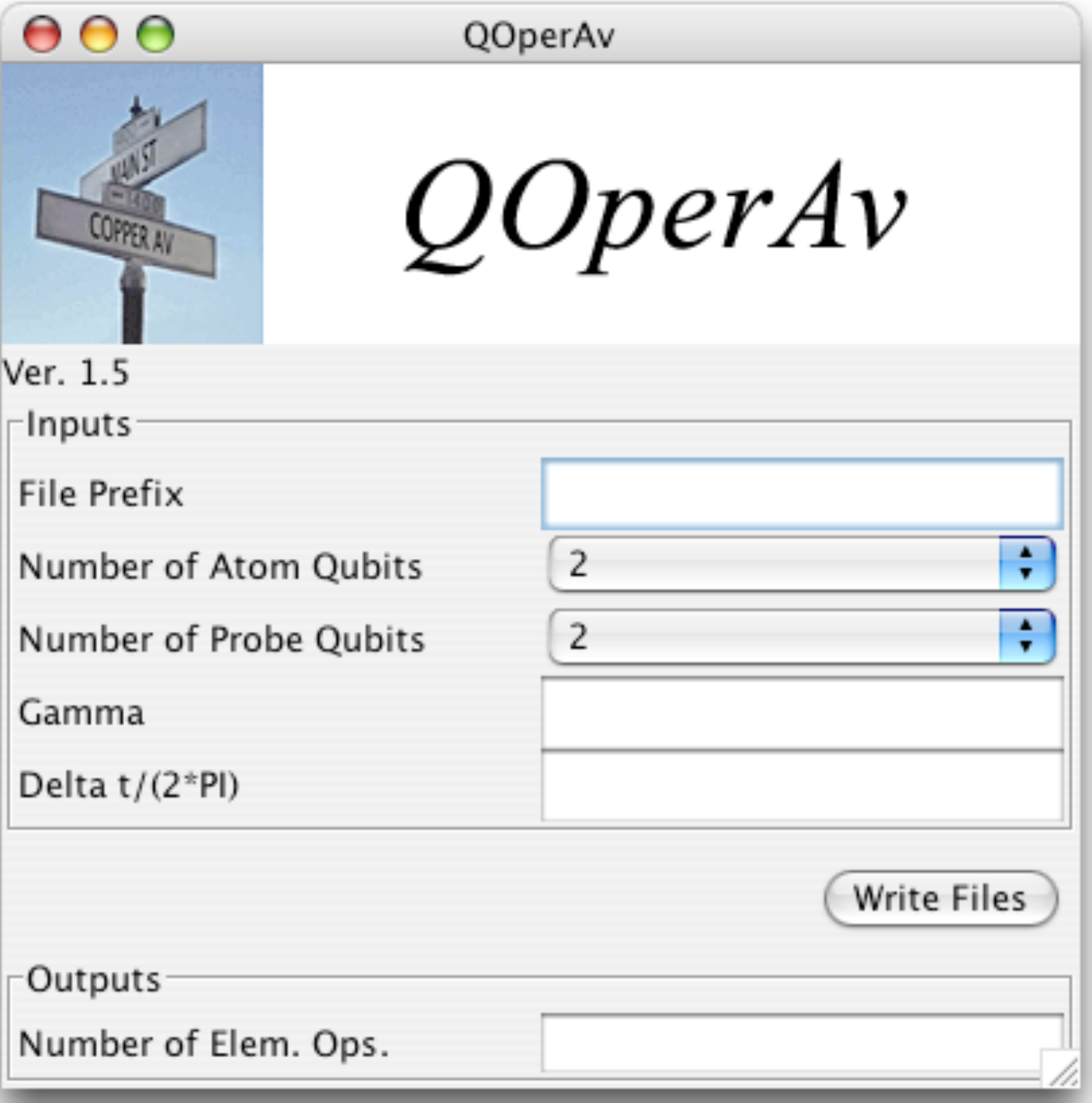}
    \caption{{\bf Control Panel} of QOperAv}
    \label{fig-qoper-av-main}
    \end{center}
\end{figure}

Fig.\ref{fig-qoper-av-main} shows the
{\bf Control Panel} for QOperAv. This is the
main and only window of QOperAv
(except for the occasional
error message window). This
window is
open if and only if QOperAv is running.
\newpage

The {\bf Control Panel}
allows the user to enter the following inputs:
\begin{description}

\item[File Prefix:] Prefix to the
3 output files
that are written when the user presses the
{\bf Write Files} button. For example,
if the user inserts {\tt test} in this text field,
the following 3 files will be written:
\begin{itemize}
\item
{\tt test\_qoa\_log.txt} This is a Log File.

\item
{\tt test\_qoa\_eng.txt}
This is an English File

\item
{\tt test\_qoa\_pic.txt}
This is a Picture File.

\end{itemize}

\item[Number of Atom Qubits:] This equals
$\nb$.

\item[Number of Probe Qubits:] This equals
$\nb_\rvj$.

\item[gamma:] This equals
$\gamma$.

\item[Delta t/(2*PI):] This equals
$\Delta t/(2\pi)$.

\end{description}

The {\bf Control Panel}
displays the following
output text boxes.

\begin{description}

\item[Number of Elementary Operations:]
This is the number of elementary operations
in the output quantum circuit.
If there are no LOOPs, this is
the number of lines in the English File, which
equals the number of lines in the
Picture File.
For a LOOP (assuming it is not nested
inside a larger LOOP), the
``{\tt LOOP k REPS:$N$}" and
``{\tt NEXT k}" lines are not counted,
whereas the lines between
``{\tt LOOP k REPS:$N$}" and
``{\tt NEXT k}"
are counted $N$ times (because
 {\tt REPS:$N$} indicates
 $N$ repetitions of the loop body).
Multiplexors expressed as a
single line are counted as a
single elementary operation
(unless, of course, they are inside a LOOP,
in which case they are
counted as many times as the loop body
is repeated).

\end{description}

\end{document}